\documentclass[twocolumn, letterpaper]{aastex61}
\usepackage{graphicx}
\usepackage[suffix=]{epstopdf}
\usepackage{natbib}
\usepackage{amsmath}
\usepackage{url}
\usepackage{xspace}

\newcommand{\Kepler}{\textsl{Kepler}\xspace}

\begin{document}

\title{Flare Activity of Wide Binary Stars with \Kepler}

\shorttitle{Wide Binary Flares}
\shortauthors{Clarke et al.}

\author{Riley W. Clarke}
\affiliation{Department of Physics \& Astronomy, Western Washington University, 516 High St., Bellingham, WA 98225, USA}
	
\author{James R. A. Davenport}
\altaffiliation{NSF Astronomy \& Astrophysics Postdoctoral Fellow}
\affiliation{Department of Physics \& Astronomy, Western Washington University, 516 High St., Bellingham, WA 98225, USA}

\author{Kevin R. Covey}
\affiliation{Department of Physics \& Astronomy, Western Washington University, 516 High St., Bellingham, WA 98225, USA}

\author{Christoph Baranec}
\affiliation{Institute for Astronomy, University of Hawai'i at M\={a}noa, Hilo, HI 96720-2700, USA}

\begin{abstract}
We present an analysis of flare activity in wide binary stars using a combination of value-added data sets from the NASA \Kepler mission. The target list contains a set of previously discovered wide binary star systems identified by proper motions in the \Kepler field. We cross-matched these systems with estimates of flare activity for $\sim$200,000 stars in the \Kepler field, allowing us to compare relative flare luminosity between stars in coeval binaries. From a sample of 184 previously known wide binaries in the \Kepler field, we find 58 with detectable flare activity in at least 1 component, 33 of which are similar in mass (q $>$ 0.8). Of these 33 equal-mass binaries, the majority display similar ($\pm 1$ dex) flare luminosity between both stars, as expected for stars of equal mass and age. However, we find two equal-mass pairs where the secondary (lower mass) star is more active than its counterpart, and two equal-mass pairs where the primary star is more active. The stellar rotation periods are also anomalously fast for stars with elevated flare activity. Pairs with discrepant rotation and activity qualitatively seem to have lower mass ratios. These outliers may be due to tidal spin-up, indicating these wide binaries could be hierarchical triple systems. We additionally present high resolution adaptive optics images for two wide binary systems to test this hypothesis. The demographics of stellar rotation and magnetic activity between stars in wide binaries may be useful indicators for discerning formation scenarios of these systems. 
\end{abstract}


\section{Introduction}

Stellar age estimation for isolated main sequence field stars is notoriously difficult. A now popular stellar dating method, known as ``gyrochronology'', is built upon a connection between stellar age and rotation made by \citet{skumanich1972}, who demonstrated that rotation decreases over time \added{for solar-like stars}, and attributed this effect to magnetic braking. Gyrochronology models use the idea that young stars of various initial periods rapidly converge to a single evolutionary track, and spin down as they shed angular momentum via magnetized winds. This rotation evolution allows observers to fit period--mass (or equivalently, period--color) isochrones to populations of stars with reliable rotation periods. However, gyrochronology models are empirically calibrated and not fully refined, resulting in intrinsic age errors of $\geq 10 \%$ \citep{meibom2015}. Furthermore, calibration of gyrochronology models necessitates the difficult task of obtaining accurate rotation measurements or independent age estimates for large populations of stars, usually from open cluster members.

\citet{skumanich1972} additionally noted a connection between rotation and chromospheric Ca HK activity, which in turn is related to the surface magnetic field strength. The common explanation for the solar-like magnetic field is dynamo generation at the interface between the rotationally decoupled core and convective envelope, or tachocline layer \citep{barnes2003}. The angular momentum loss described by \citet{skumanich1972} is driven by magnetized winds in a process known as magnetic braking. This rough trend between stellar rotation and magnetic activity was first quantified by \citet{pallavicini1981}, who showed that X-ray luminosity, a tracer of magnetic activity, scaled as $L_X\;\alpha\;(v\sin i)^{1.9}$.

 Like X-rays, flares are tracers of magnetic activity, arising from magnetic reconnection events on the stellar surface \citep{cram&mullan1979}. Observation of a broken power law decay in flare activity vs. rotation period was made by \citet{davenport2016} for \Kepler stars with spectral types later than G8. Below a critical Rossby number ($Ro_{sat} = 0.036\pm0.004$), the magnetic dynamo becomes saturated and tracers of magnetic activity remain constant as the period shortens. The slope of the power law decay in the non-saturated regime is consistent with that of other tracers of magnetic activity, e.g. $H\alpha$ or X-ray emissions. This result demonstrated the viability of flares as tracers of magnetic activity, however the evolution of flare rates in the age--activity paradigm remains uncertain.

Wide binaries are ideal laboratories for testing the age--rotation--activity paradigm. Wide binaries are observed as pairs of common proper motion stars with separations ranging from $\sim$1000 to $\sim$10,000 AU \citep{moe&stefano2017}. Though they are weakly bound gravitationally, the odds of tidal capture is low, and thus it is widely thought that these binaries formed from the same progenitor molecular cloud. Indeed, \citet{andrews2017(1)} showed consistent metallicity and elemental abundances in wide binaries in the Tycho-Gaia Astrometric Solution (TGAS), further confirming the common origin of wide binary stars. This allows us to make flare activity comparisons between coeval, separately resolved binary components. If flare activity evolves coherently with rotation as the star ages and loses angular momentum, we expect to observe similar flare signatures between stars of similar mass and age. Wide binaries have previously been used as coeval laboratories to compare magnetic activity by \citet{gunning2014}, who compared chromospheric $H\alpha$ activity between coeval M-dwarf twins. In a similar fashion, by comparing flare activity levels from known wide binaries in the \Kepler field, we can study the universality of flares as a tracer of magnetic dynamo evolution. 

In this paper, we test this paradigm with a unique comparison of flare activity between wide binary components in the \Kepler field. In \S\ref{sec:sample}, we describe the selection process of our sample. In \S\ref{sec:analysis}, we identify outlier populations in the flare activity comparison space and we explore them by examining mass ratios, physical separations, and rotation periods of the stars in these pairs. In \S\ref{sec:triples} we propose a potential cause of the anomalously elevated flare activity in the outlier populations in this section, and present Adaptive Optics imaging for two targets in the outlier populations. Lastly, in \S\ref{sec:summary} we summarize the key results of this work.

\section{Sample Selection}
\label{sec:sample}

This section describes the constituent data sets that make up our final sample, as well as follow-up observations made examine unexpected behavior in two outlier systems.

\subsection{Wide Binary Data}

Our flare sample comes from the intersection of two datasets, the first of which is the catalog of flare events identified in \Kepler light curves by \citet{davenport2016}. In this study, Davenport searched for flares from all 207,617 stars in the \Kepler MAST archive up to and including Data Release 24. The Davenport catalog contains estimates of the relative flare luminosity ($L_{fl}/L_{kp}$) for each star. The $L_{fl}/L_{kp}$ metric is defined as the integrated flux from all identified flares, divided by the total flux over the full monitoring period, and is used to quantify overall flare activity. This measures the relative luminosity produced by flares compared with the luminosity emitted in the \Kepler bandpass for each star. The $L_{fl}/L_{kp}$ metric was shown by \citet{lurie2015} to be an effective means of comparing flare activity levels for stars with similar spectral types and distances, such as found in many wide binary systems. \citet{davenport2016} measured the relative flare luminosity using all events detected as part of an automated survey of flares from all \Kepler light curves. For each star, artificial flare injection and recovery tests were used to determine the noise floor for flare recovery. Though the published Davenport sample only included stars with large numbers of high-probability flare events, the survey was run for all available \Kepler stars (Davenport in prep).

Wide binaries in the \Kepler field were sourced from a list of 184 pairs with measured rotation periods compiled by \citet{janes2017}. These wide binaries were vetted by identifying co-moving stars using their proper motions, although \citet{janes2017} notes that up to 15\% of these binaries may be false positives. This data set included measured angular separations and rotation periods for these pairs, however does not include radial velocity measurements for these pairs. The intersection of these data sets provided $L_{fl}/L_{kp}$ estimates, masses, and rotation periods for 184 verified wide binaries in the \Kepler field.

\begin{figure}[!ht]
\centering
\includegraphics[width=0.45\textwidth]{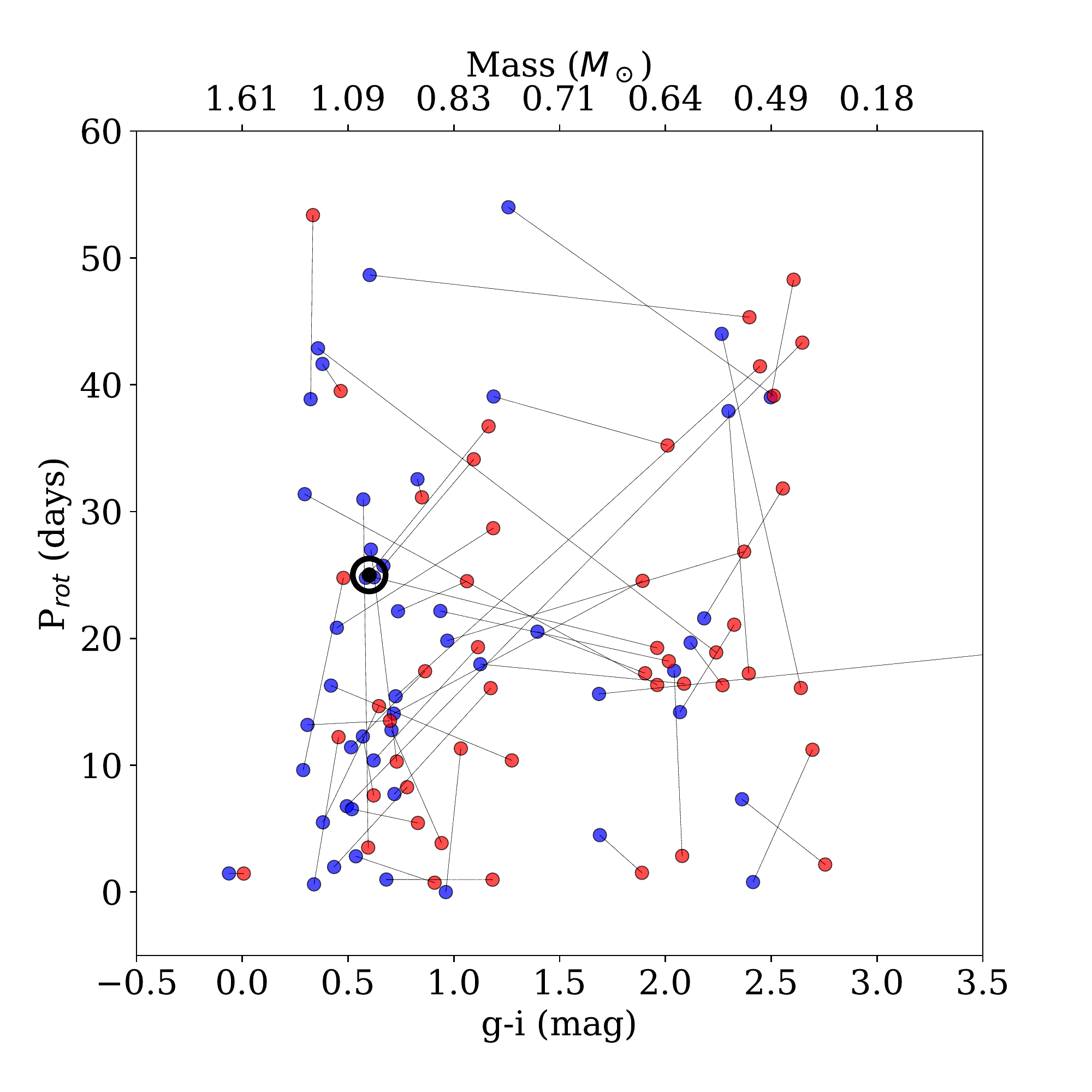}
\caption{98 stars from this sample plotted in mass-period space. Stars are connected by lines to their binary counterpart. 18 stars in our sample had no available $g-i$ color in the Kepler Input Catalog. Primary (higher mass) components are labeled blue, secondary (lower mass) components are labeled red. This figure is similar to Figure 11 from \citet{janes2017}, which displays common proper motion pairs in period--color space. The Sun has been added for reference.
}
\label{fig:mass-prot}
\end{figure}

Additional filtering was performed to ensure the activity from at least one component in each pair was high enough to be distinguishable from background noise. Our final sample contained 58 pairs with at least one component satisfying a relative flare luminosity criteria of $L_{fl}/L_{kp} > 10^{-7}$, and 49 systems that satisfy this criteria in both components. Table \ref{table} contains properties for all 116 stars in the final sample. To characterize this final sample, Figure \ref{fig:mass-prot} displays each star in color (mass)--period space, connected to its binary counterpart. 
Here we use the $g-i$ color adopted in \citet{davenport2016} as a proxy for mass, as it has been used previously to track the main-sequence stellar locus for FGKM stars \citep[e.g.][]{covey2007}.

\subsection{Adaptive Optics Imaging}
\label{sec:AO}

In addition to the wide binary sample, we also obtained follow-up Adaptive Optics observations for two systems in our sample. We used the NIRC2 narrow-field infrared camera mounted on the Keck II Adaptive Optics (AO) system. The NIRC2 camera was operated in 9.9 mas/pixel mode, resulting in a $\sim$10" field of view. The observations took place on 2017 Aug 3, 2017 Aug 9, and 2017 Aug 10. The filters, integration times, and number of dithers for each target were as follows: 
\begin{itemize}
    \item KIC 7871442: $Kp$, 4 dithers with 5-10 second integrations
    \item KIC 7871438: $Kp$, 3 dithers with 5 second integrations
    \item KIC 10536761: $K$, 6 dithers with 2 second integrations
    \item KIC 8888573: $K$, 4 dithers with 1-2 second integrations
\end{itemize}
For each of the four targets, the dithered images were bias and flat corrected, and then manually aligned and median-combined. These data are discussed further in \S\ref{sec:triples}.

\section{Analysis}
\label{sec:analysis}

With our sample of 58 flaring wide binary pairs, we compared flare activity between the components, with the expectation that coeval stars of similar mass should display 1-to-1 flare activity levels. After observing large scatter in the relative flare luminosity between A- and B-components, we sort the sample into three distinct populations.

\subsection{Flare Activity Comparison}

\begin{figure}[!ht]
\centering
\includegraphics[width=0.45\textwidth]{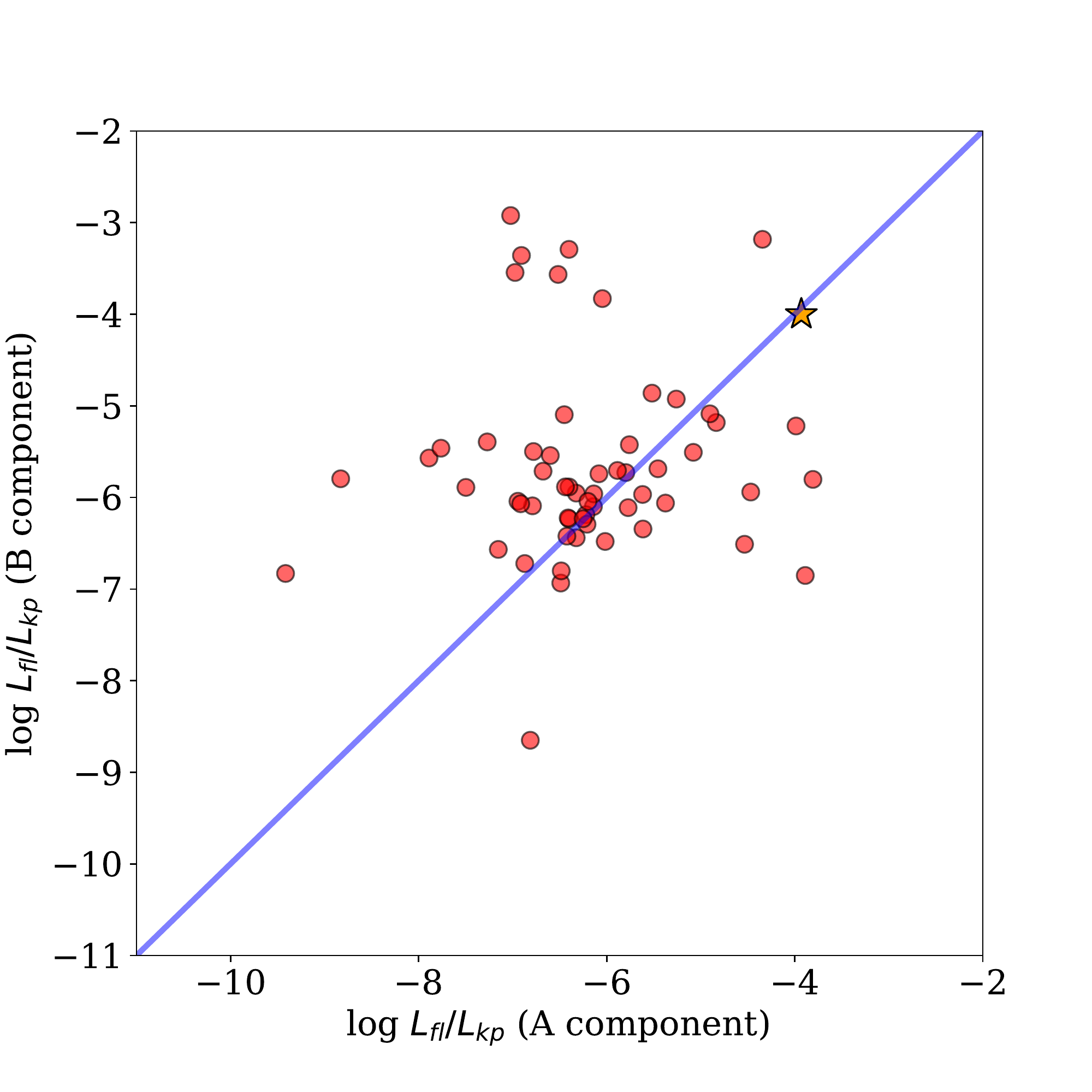}
\caption{Component-wise $L_{fl}/L_{kp}$ comparison for 58 wide binary systems with detected flare activity. The central blue line is the expected relationship for coeval, equal mass components. The M5 + M5 binary, GJ 1245AB, from \citet{lurie2015} is added for comparison (orange star). While not in our sample, GJ 1245AB represents an ideal example of coeval binary components with equal and detectable flare rates.
}
\label{fig:AB1}
\end{figure}

The combination of data sets described in \S\ref{sec:sample} allows us to directly compare flare activity in coeval wide binary components. Figure \ref{fig:AB1} compares the relative flare luminosity of the primary star (A-component) vs. that of the of the secondary star (B-component) for the 58 pairs in our sample. Here we use $L_{fl}/L_{kp}$, defined in \citet{lurie2015} and \citet{davenport2016}, as the relative flare luminosity of a single star. In the age--rotation--activity paradigm, we expect coeval, equal-mass stars to exhibit similar flare luminosities, thus following the 1-to-1 line in Figure \ref{fig:AB1}. For coeval stars with lower mass ratios, the secondary is expected to have higher activity \citep[e.g.][]{west2008,douglas2014}, and therefore fall above the 1-to-1 line in Figure \ref{fig:AB1}. While many wide binary pairs appear to satisfy the expected age--activity relationship, some do not. Note, no comparable white light flare luminosity data is available for the Sun, and thus we cannot yet add it to Figure \ref{fig:AB1}.

\begin{figure}[!ht]
\includegraphics[width=0.45\textwidth]{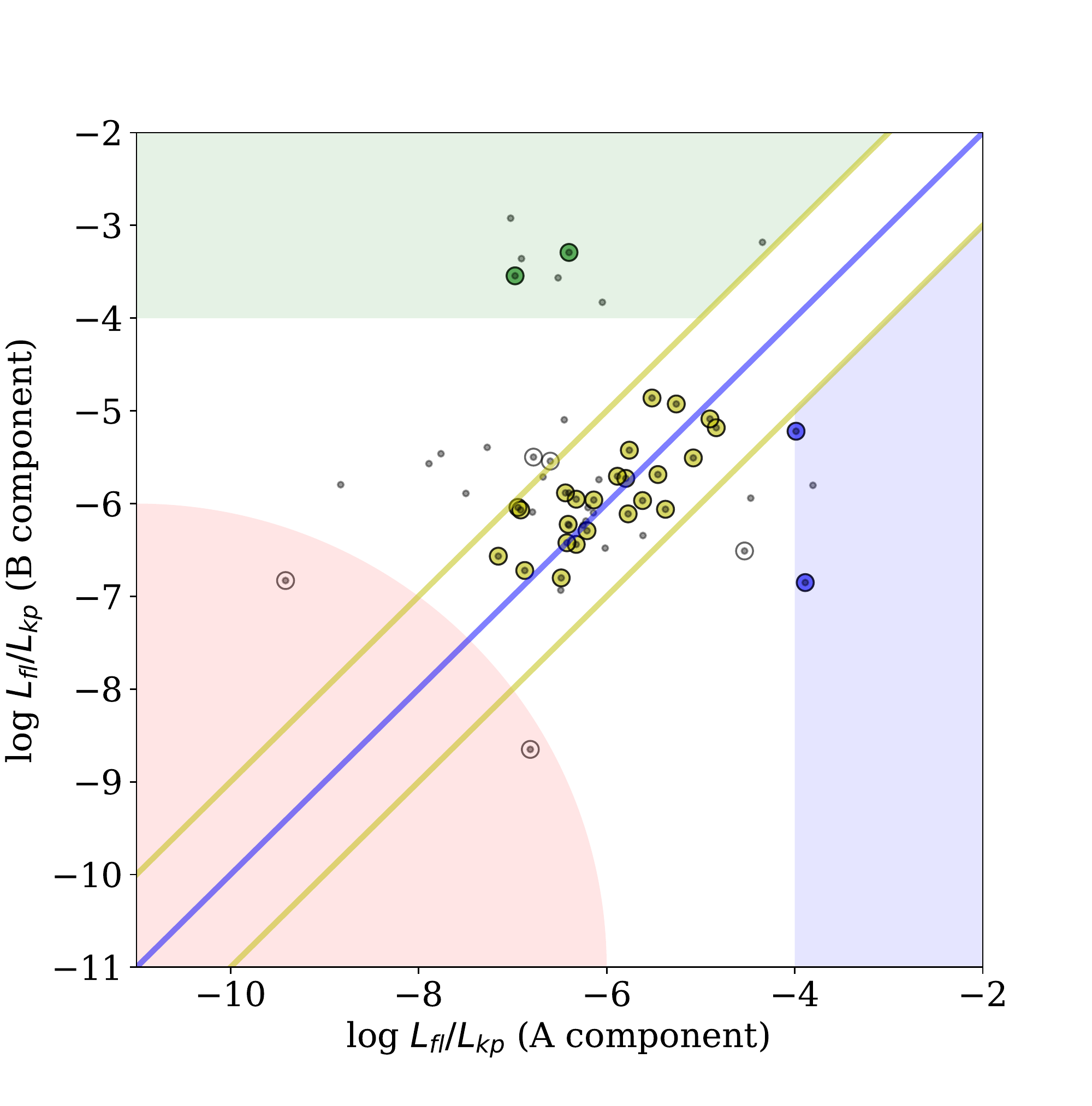}
\caption{Component-wise $L_{fl}/L_{kp}$ comparison for 58 wide binaries with detected flare activity. The central blue line is the expected relationship for equal mass coeval components. The area between the yellow lines contains pairs whose difference in flare luminosity may plausibly be explained by a solar-like activity cycle. The green region contains pairs where the secondary star is significantly more active, while the blue region contains pairs where the primary is more active. The red region denotes the minimum flare detection threshold, below which flare signals are indistinguishable from noise. Circled points are pairs with q $\geq$ 0.8, uncircled points are pairs with q $<$ 0.8.
}
\label{fig:highlight}
\end{figure}

To better understand the activity scatter in Figure \ref{fig:AB1} we have highlighted several important regions of this parameter space in Figure \ref{fig:highlight}. Our sample was generated in \S\ref{sec:sample} by requiring at least one component of each binary have a relative flare luminosity from \citet{davenport2016} of $\log(L_{fl}/L_{kp}) > -7$. However, since \citet{davenport2016} only provides statistical uncertainties that are too small to be useful for our analysis, we opted to further cull our sample of any systems where both components had $\log(L_{fl}/L_{kp}) < -6$, eliminating two binaries in the red region of Figure \ref{fig:highlight}.

While episodic outbursts of flaring can explain modest deviation from the 1-to-1 line in Figure \ref{fig:highlight}, they do not account for the significant deviation of pairs in the green and blue regions. Solar-like activity cycles are known to exist on other stars with outer convective envelopes \citep[e.g.][]{see2016}, which we expect to cause intrinsic scatter about the one-to-one line of flare activity for normal, equal-mass stars in this space. The only robust measurement of the variation in flare rate over the course of a magnetic activity cycle comes from the Sun, where the flare rate has been observed to vary by approximately an order of magnitude \citep{veronig2002}. We therefore conservatively classified binary pairs whose flare activity was within a $\pm$1 dex region of the one-to-one line as being consistent with normal activity cycles. This selected a sample of 48 binary pairs where the flare activity from both stars is consistent with a single age--activity evolution for stars of equal mass. The presence of a majority of such systems in our sample is a strong validation of flares as a valuable tracer of stellar magnetic activity.

However, many binaries in our sample do not have equal masses, and lower mass stars are known to have longer magnetic activity lifetimes \citep{west2008}. Therefore, in Figure \ref{fig:highlight} we have circled 33 pairs with mass ratios $>$ 0.8 to emphasize that: a) Most pairs in the yellow region are roughly equal mass binaries and b) Some pairs in the outlier regions have mass ratios near 1, so their elevated flare activity cannot be explained by a disparity in mass. For binaries with mass ratios less of $q<1$, we would generally expect the secondary component to have a higher relative flare luminosity at any age, and as a result to cause points to scatter above the line of unity in Figure \ref{fig:highlight}. We will explore the potential impact of stellar mass ratios further in \S\ref{sec:triples}, but note that seven systems in Figure \ref{fig:highlight} had relative flare luminosities with the B-component modestly higher than the A-component, consistent with this model.

\begin{figure*}[ht]
        \centering
  	\includegraphics[width=3in]{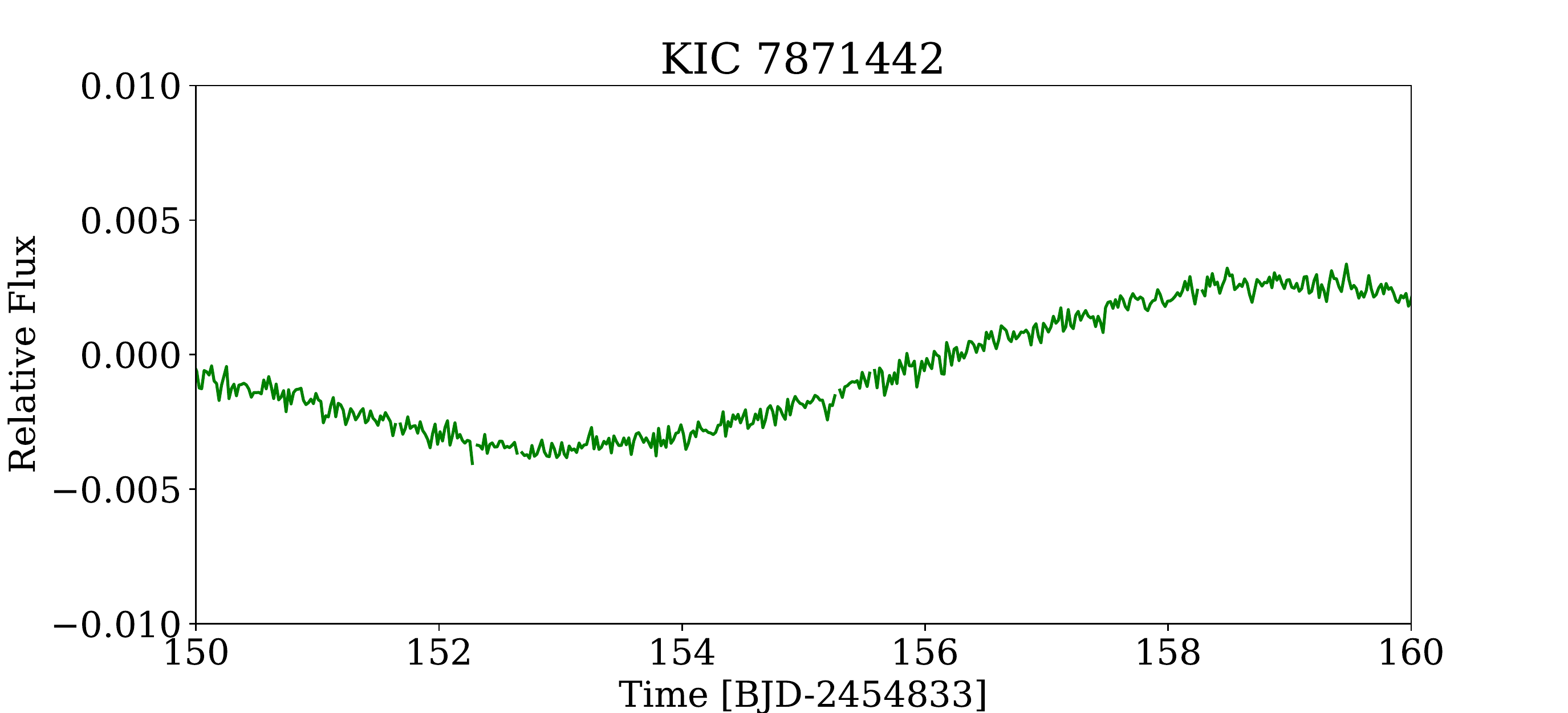}
	\includegraphics[width=3in]{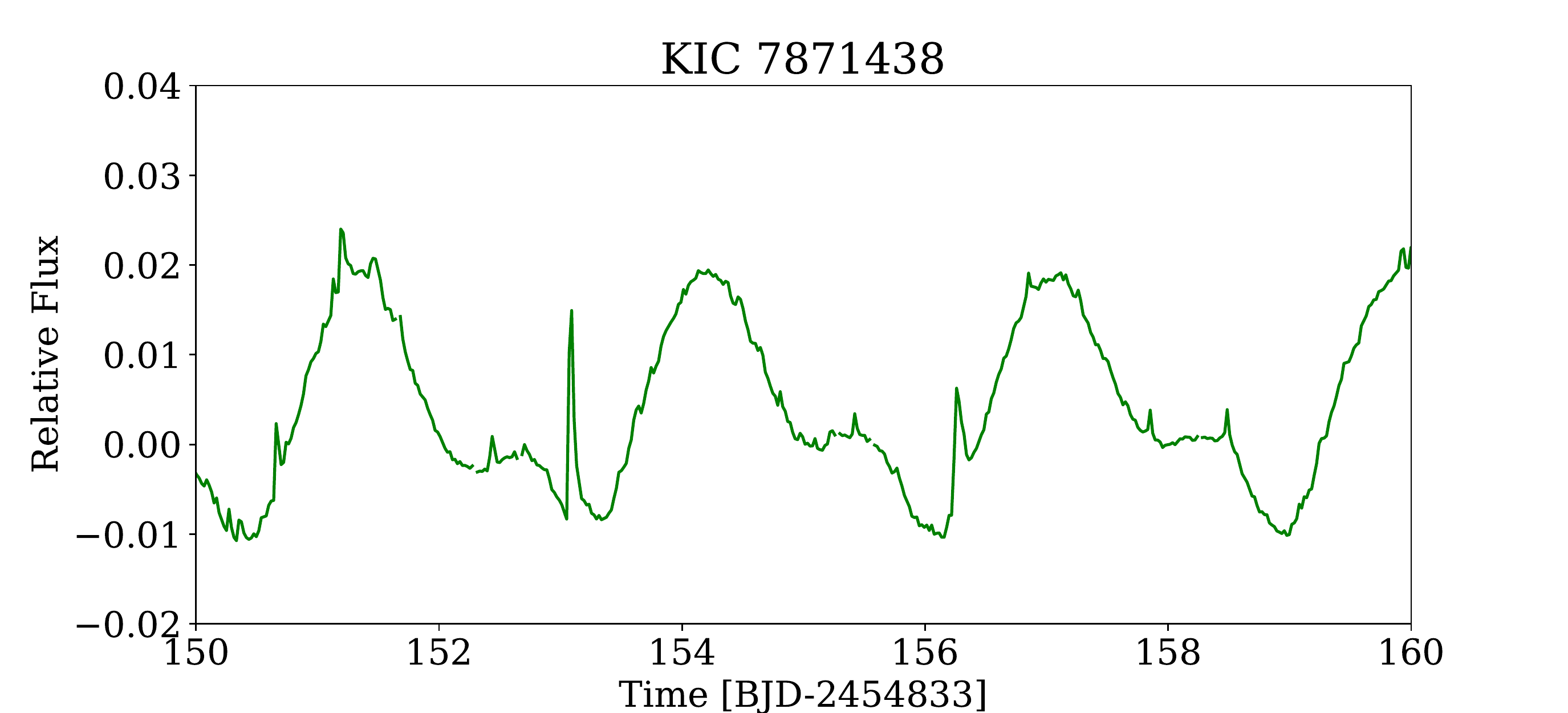}
	\includegraphics[width=3in]{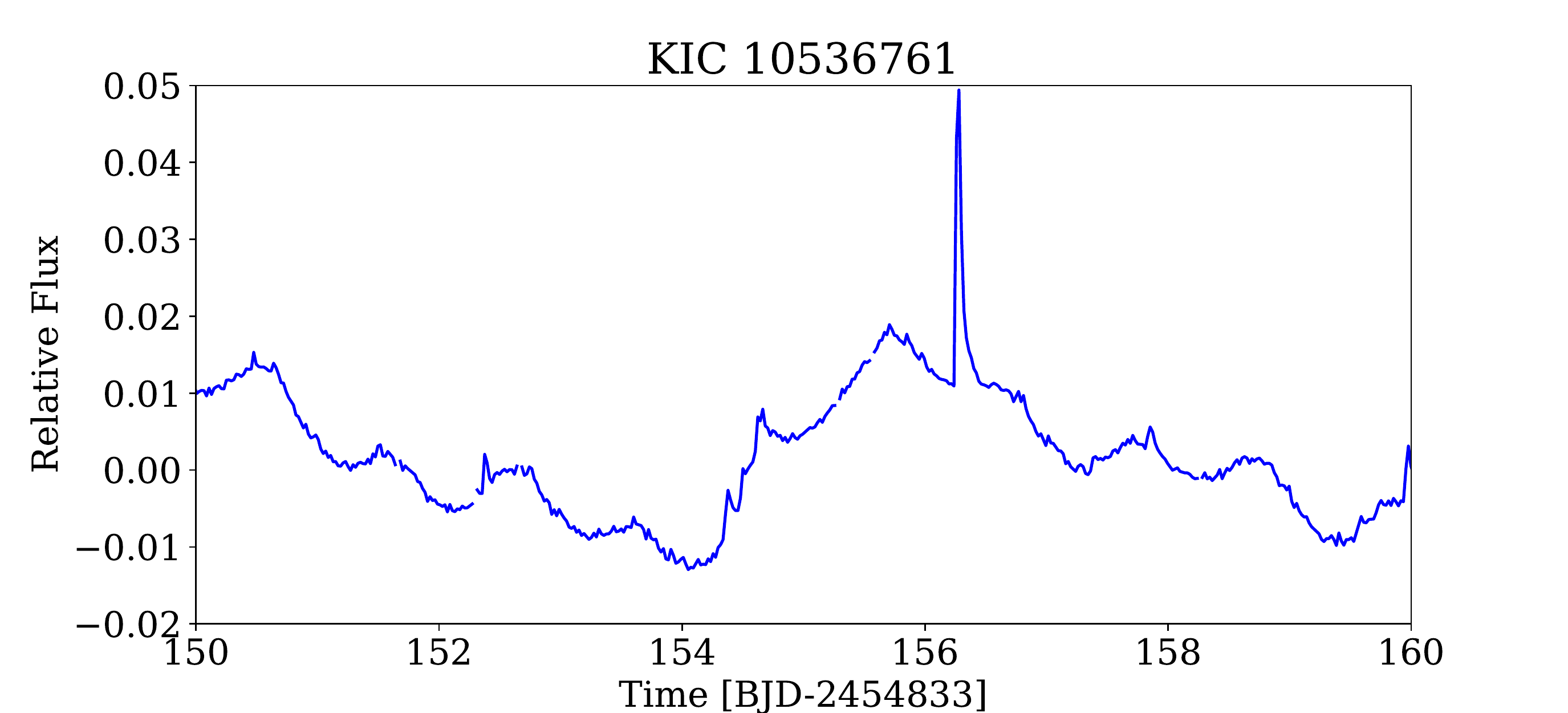}
	\includegraphics[width=3in]{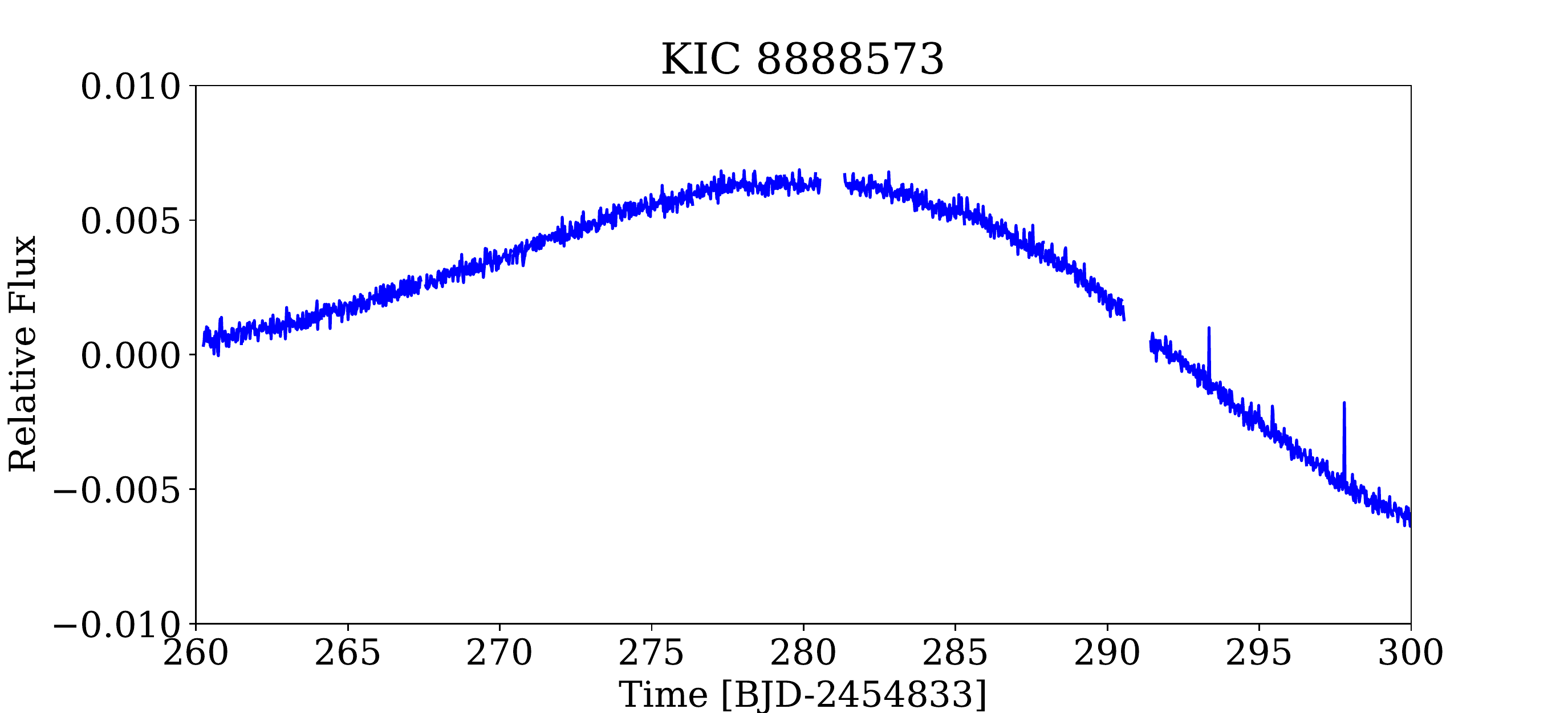}
        \caption{ Light curves for two wide binary systems from the ``green'' and ``blue'' populations in Figure \ref{fig:highlight}. The A- \& B-components are labeled below each panel. 10-day sections of the light curves we chosen to illustrate the varying flare and rotation signals, except KIC 8888573, whose slow rotation required 40 days to display. These examples were selected due to having mass ratios near 1.} 
    \label{fig:lc}
\end{figure*}

Interestingly, a small number of wide binary systems exhibit a significant asymmetry in the relative flare luminosity between the A- and B-components. These have been selected in Figure \ref{fig:highlight} as systems where only one component would be considered active in \citet{davenport2016}, i.e. having $\log(L_{fl}/L_{kp}) > -4$ for only one stellar component. A total of 10 systems met this criteria, with 7 having significantly higher activity in the B-component, and 3 in the A-component. We call these systems ``outliers'' throughout the rest of this analysis, as they violate multiple expectations for the age--activity paradigm for coeval stars. Light curves for two examples of these outlier systems are shown in Figure \ref{fig:lc}. While the flare activity asymmetry of the green points (B-component more active) in Figure \ref{fig:highlight} is larger than we would predict, the asymmetry of the blue points (A-component more active) are completely unexpected, and are an important challenge to the formation and evolutionary theory of wide binary systems.

The presence of the ``outlier'' systems found in Figure \ref{fig:highlight}, i.e. those with vastly different flare activity levels between binary components, is an unexpected result of this study. We note one possible explanation is that these systems are not true wide binaries, and instead chance alignments of stars with similar proper motions. Indeed, the \citet{janes2017} catalog estimates a contamination rate of $\sim$15\%, similar to the occurrence rate of stars in our ``green'' and ``blue'' regions. However, \citet{janes2017} also produced a probability metric for chance association of proper motions given the local density of stars in the  Galactic field. In Figure \ref{fig:janesprob} we show the distribution of this probability metric for the ``normal'' (yellow) and ``outlier'' (blue and green) systems. The outliers are not preferentially more likely to be contaminated by associated stars, and we therefore assume this population is astrophysical.

\subsection{Possibility of Chance Alignments}

\begin{figure}[!ht]
\centering
\includegraphics[width=0.45\textwidth]{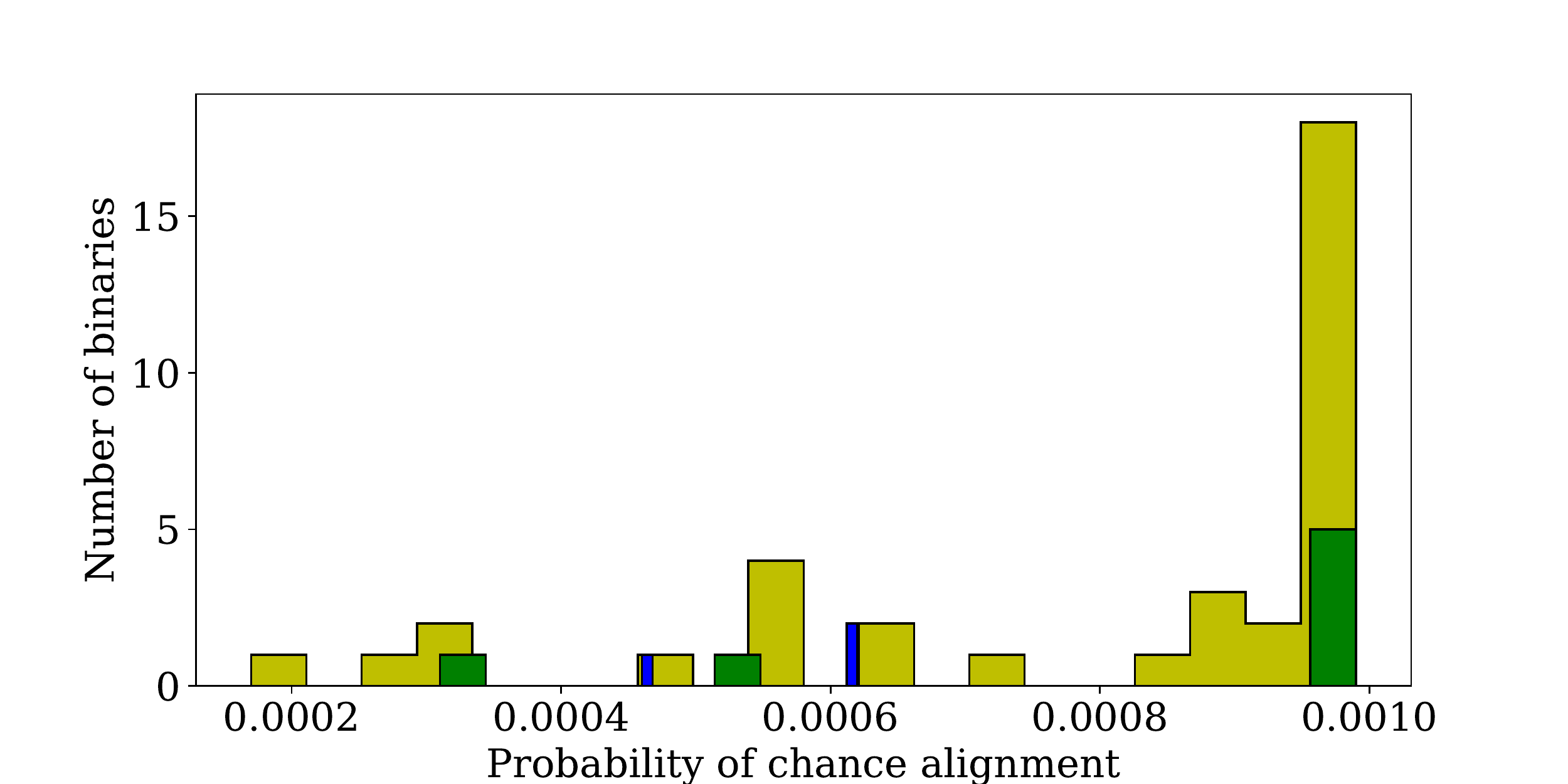}
\caption{Distribution of chance alignment probabilities computed by \citet{janes2017}, for the ``normal'' (yellow) and ``outlier'' (green and blue) activity configurations defined in Figure \ref{fig:highlight}. No significant difference in this probability is seen between the normal and outlier systems.
}
\label{fig:janesprob}
\end{figure}

The presence of the ``outlier'' systems found in Figure \ref{fig:highlight}, i.e. those with vastly different flare activity levels between binary components, is an unexpected result of this study. We note one possible explanation is that these systems are not true wide binaries, and instead chance alignments of stars with similar proper motions. Indeed, the \citet{janes2017} catalog estimates a contamination rate of $\sim$15\%, similar to the occurrence rate of stars in our ``green'' and ``blue'' regions. However, \citet{janes2017} also produced a probability metric for chance association of proper motions given the local density of stars in the  Galactic field. In Figure \ref{fig:janesprob} we show the distribution of this probability metric for the ``normal'' (yellow) and ``outlier'' (blue and green) systems. A two-sample K-S test of the outlier probability distribution versus the normal population gave a p-value of 0.83, indicating that the outliers are not preferentially more likely to be contaminated by associated stars. 

\begin{figure}[!t]
\centering
\includegraphics[width=0.45\textwidth]{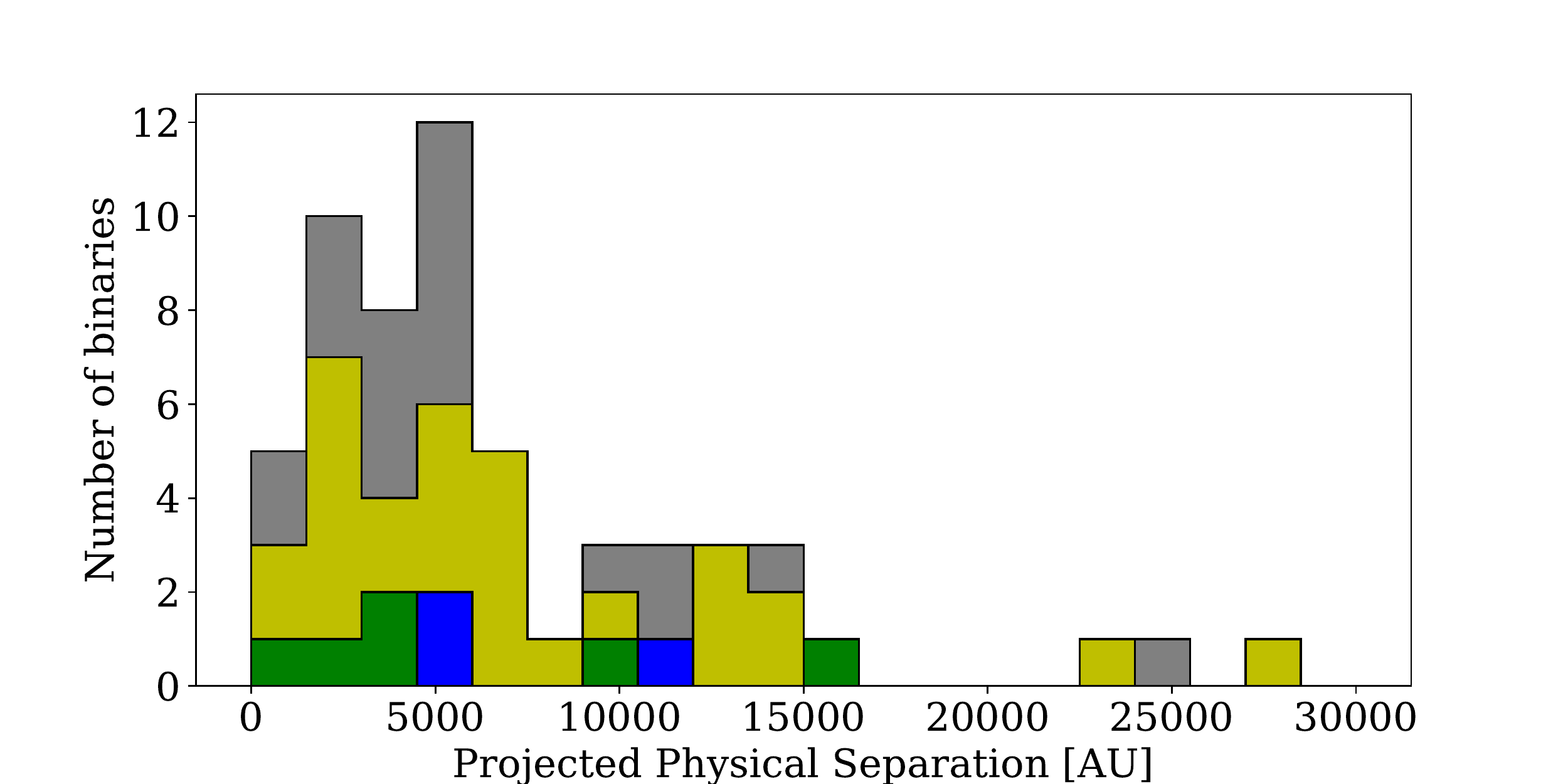}
\caption{Distribution of projected physical separations between wide binary components for each activity population defined in Figure \ref{fig:highlight}. No clear differences in separation distributions can be seen for the three different configurations. 
}
\label{fig:sepdist}
\end{figure}

It has previously been suggested that wide binaries may form via multiple separation scenarios, such as three body dynamics, radial migrations, interactions with cluster members or turbulent fragmentation \citep{lee2017}. These different formation channels may result in a range of system separations. \citet{dhital2015} suggest the presence of two separation populations in SLoWPoKES-II sample, which might indicate different formation or dynamical histories. However, recent work by \citet{andrews2017} has detected no evidence of bimodality for projected separations $s \lesssim 1$ pc in the Tycho-\textit{Gaia} astrometric solution (TGAS) catalogue. Figure \ref{fig:sepdist} displays the distribution of projected physical separations in AU for all three activity populations. No noticeable difference in physical separation is seen between the different activity populations. Though our sample is small, we see no evidence of a bimodal or broken distribution. This suggests that whatever mechanism is responsible for the observed activity asymmetry of the outlier systems does not affect the present day physical separation of the system, or does not affect it enough to detect in such a small sample of systems.

To better understand the origin of these unusual systems, we explore various properties of the binaries, including mass ratio and rotation periods of the constituent stars.

\subsection{The Impact of Mass Ratio}

\begin{figure}[!t]
\centering
\includegraphics[width=0.45\textwidth]{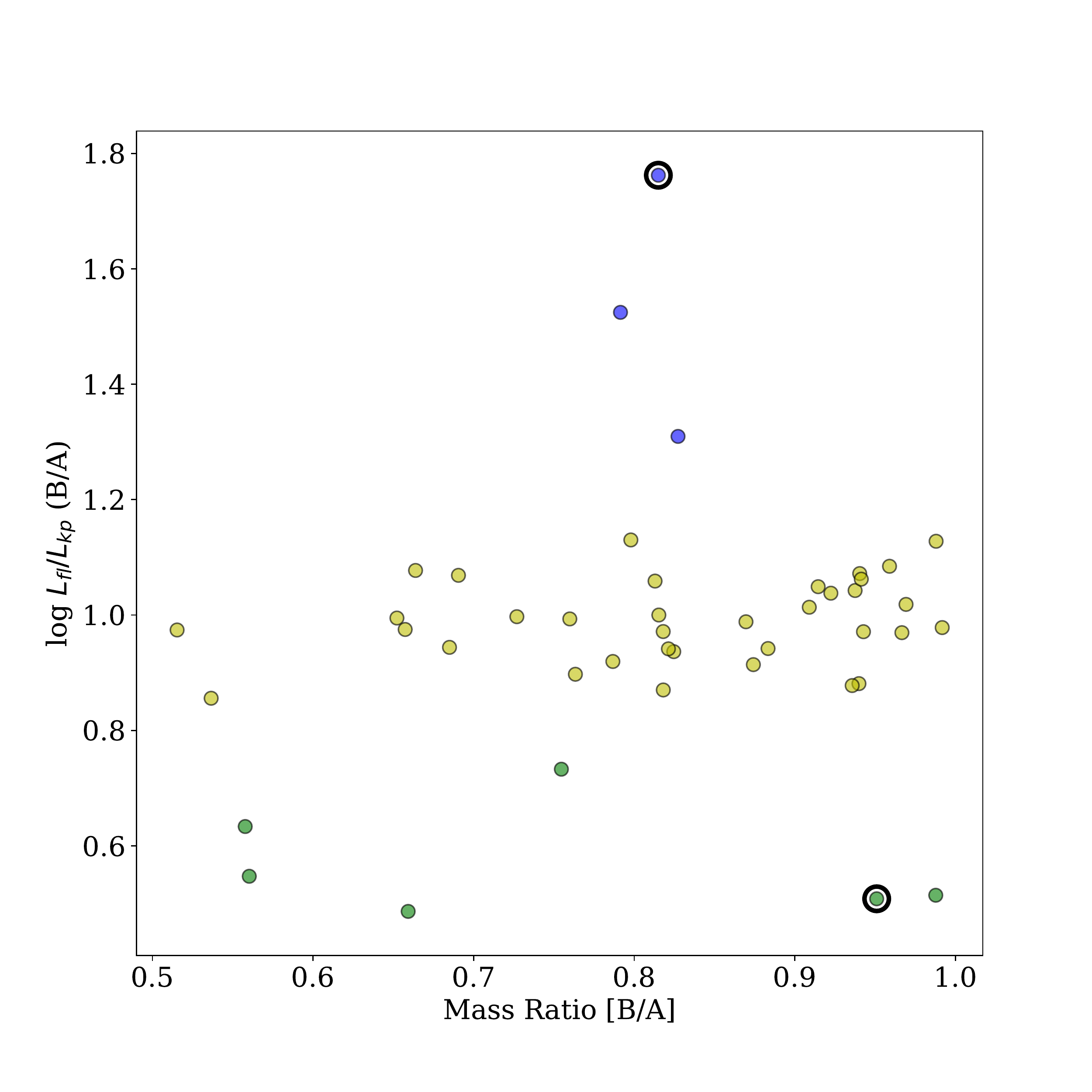}
\caption{Mass ratio vs. $L_{fl}/L_{kp}$ ratio for the three populations shown in Figure \ref{fig:highlight}. The two pairs shown in Figure \ref{fig:lc}, KIC 7871442 -- KIC 7871438 (green), and KIC 10536761 -- KIC 8888573 (blue) are highlighted (bold black circles).
}
\label{fig:massplot}
\end{figure}

\begin{figure}[!t]
\centering
\includegraphics[width=0.45\textwidth]{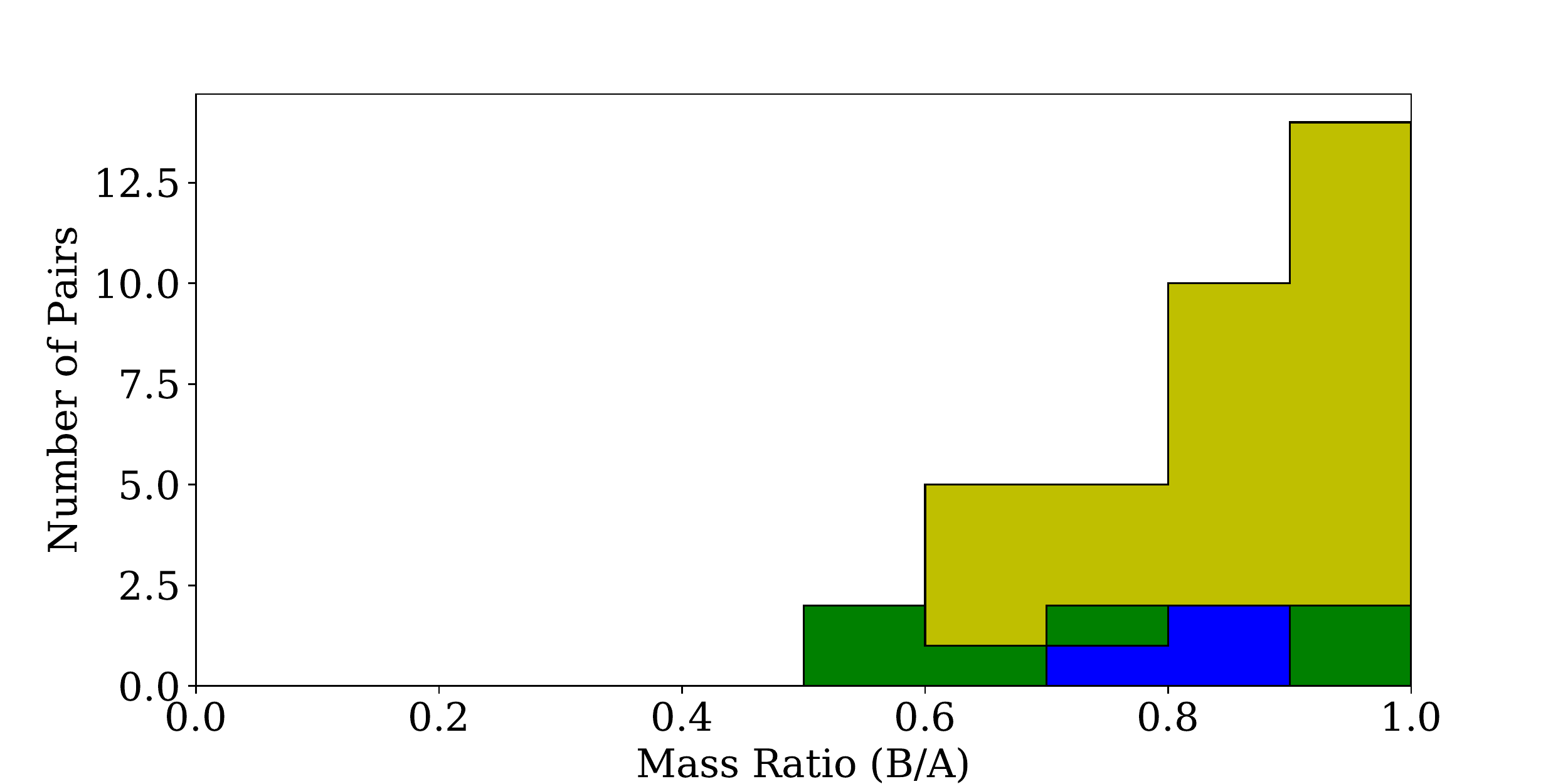}
\caption{Distribution of binary mass ratios for each activity configuration. The apparent uniformity of the green and blue distributions raises doubts as to whether mass ratios can be considered the sole reason for the activity asymmetry seen in Figure \ref{fig:highlight}.
}
\label{fig:massdist}
\end{figure}

Stars of different masses develop along different evolutionary tracks, adding a parameter degeneracy to the age--rotation--activity relation. Lower-mass stars are more active and have longer active lifetimes, resulting in a wide binary with a low mass ratio displaying elevated flare activity in the secondary component. To examine this potential cause of the green \& blue populations seen in Figure \ref{fig:highlight}, the mass ratio and $L_{fl}/L_{kp}$ ratio is shown for each system in our sample in Figure \ref{fig:massplot} and the distribution of mass ratios for each population is shown in Figure \ref{fig:massdist}. The distribution of systems in this space is unexpected given the fact that low mass companions are expected to display enhanced activity. While many binaries in the green and yellow populations display low mass ratios, two systems in the green activity bin have mass ratios $>$ 0.9. These systems defy the expectation that same-mass and same-age stars should display similar flare rates. For example, the pair highlighted in Figure \ref{fig:massplot} contains two K dwarfs with a mass ratio of 0.98, but one of the stars, KIC 7871438, has a flare luminosity ratio over 3 dex greater than its companion, KIC 7871442.


Using the flare frequency evolution model of Davenport et al. (2017 in prep), we created a model of the relative flare luminosity evolution for binary stars. We explored mass ratios in the range $0.5<q<1$, consistent with those found in our sample in Figure \ref{fig:massdist}. Flare rates were calculated for ages between 10 Myr and 1 Gyr, and the flare frequency distribution was integrated for six orders of magnitude in energy for both stars. As expected, the lower-mass components demonstrated higher relative flare luminosities, and the systems evolved along nearly straight lines in the space of Figure \ref{fig:highlight}. For the lowest mass ratio systems in our sample, $q=0.5$, the B-component had relative flare luminosities of $\sim$1.5 dex above that of the A-component. This model suggests that the binary mass ratios can explain the modestly higher B-component systems above the one-to-one line in \ref{fig:highlight}, but not the outliers in green, nor those in the blue with the more active A-component. 

To examine if low mass ratio systems are over-represented in the outlier populations, we computed a K-S test on the yellow versus the blue and green populations together. This resulted in a p-value of 0.2678, indicating that we cannot rule out the null hypothesis that the green/blue and yellow samples are drawn from the same distributions. However, given the small sample size, this K-S test cannot prove that the green and blue populations are not uniformly distributed.

\subsection{The Rotation -- Activity Connection}

\begin{figure}[!ht]
\centering
\includegraphics[width=0.45\textwidth]{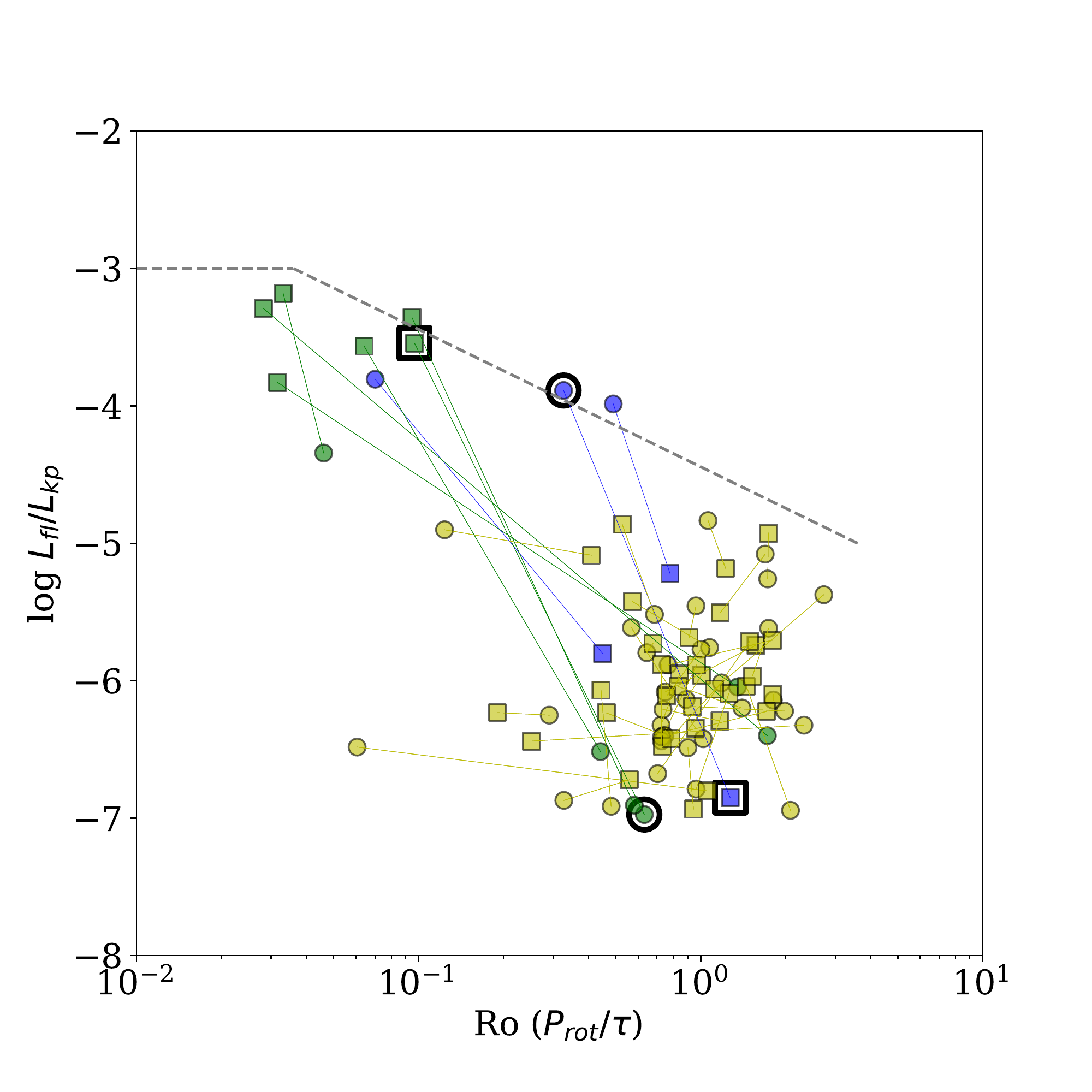}
\caption{Rossby number vs. activity for 116 individual stars from the 58 pairs in Figures \ref{fig:AB1} \& \ref{fig:highlight}. As in Figure \ref{fig:highlight}, yellow points are stars in pairs with similar flare activity, green points are stars in pairs where the B-component is more active and blue points are stars in pairs where the A-component is more active. The points are connected by lines to their binary companions. A-components are represented by circular points, while B-components are presented by square points. The two pairs whose light curves are shown in Figure \ref{fig:lc}, KIC 7871442 -- KIC 7871438 (green), and KIC 10536761 -- KIC 8888573 (blue) are highlighted (bold black lines). The grey dashed line is the broken power law decay from \citet{davenport2016}. 
}
\label{fig:rotation}
\end{figure}

While the mass ratio could explain a spread in activity for some of the wide binaries in our sample, many pairs in the asymmetric activity populations have mass ratios near one. This opens the possibility that rotationally-driven surface magnetic activity is different in these pairs. Rotation periods for all stars in our sample were provided by \citet{janes2017} via the autocorrelation function. In Figure \ref{fig:rotation}, Rossby number ($P_{rot}/\tau$) versus $L_{fl}/L_{kp}$ are shown for components in each of the activity populations described in Figure \ref{fig:highlight}. Rossby number is a dimensionless parameter used to normalize rotation periods in variable-mass stellar populations. It is defined as $P_{rot}$ in days divided by $\tau$, the convective turnover timescale. The overall trend between Rossby number and flare activity seen in Figure \ref{fig:rotation} for all stars generally follows the rotation--flare connection found by \citet{davenport2016}, serving as a further validation of flares tracing the rotation-generated magnetic activity. In this space we find that for outliers, the more active component always has a smaller Rossby number (faster rotation). This observation indicates that rotation, not mass, is the driving factor of the pairs in the green and blue populations. For pairs with mass ratios near 1, rotation periods are the only property that can explain the wide scatter in flare activity seen in Figure \ref{fig:highlight}. Indeed, \citet{janes2017} noted many wide binaries where gyrochronology relations gave disparate ages between A- \& B-component. The overall trend between rotation and activity in this space suggests that these systems are individually consistent with a single rotation--activity relation, but may not conform to a single age--rotation relation.

\section{Hierarchical Triples \& the Effects of Tides}
\label{sec:triples}

The rotation and activity asymmetry seen in a fraction of our equal-mass outlier binaries indicates that the spin of one star in these systems has been perturbed from the expected rotation evolution and angular momentum loss. The most surprising feature of our results is the existence of objects in the blue population in Figure \ref{fig:AB1}, where the higher-mass star appears to be rotating anomalously fast, as we would naively expect the lower-mass star to be more sensitive to tidal forces. These outlier systems may therefore be due to tidal spin-up by an unresolved third companion around the affected star. Dynamical three-body models \citep[e.g.][]{toonen2016} have shown that hierarchical triple systems display complex behavior such as Lidov-Kozai cycles that give rise to enhanced tidal effects, which could alter the observed rotation rate. \citet{dhital2015} describes multiple formation channels for wide binary systems that result in high fractions of triple and quadruple star systems, although they note that no particular pathway appears to be dominant. We further believe the likelihood of our outlier systems containing many third bodies is plausible because a significant fraction of low-mass stars ($\sim10\%$) are in the field are triple or higher multiplicity systems \citep{tokovinin2008}. 

Besides tidal spin-up from a third body, we have considered two alternative explanations for the anomalous rotation periods observed in this work. The first is that these outlier systems are pairs of {\it young stars} that have yet to converge to a single mass-rotation evolutionary track, such as those described in \citet{agueros2011}. These rotation evolution tracks show that while young, low-mass stars may have a large spread in the $P_{rot}$-mass space, and they will eventually converge to a single evolutionary track useful for gyrochronology. These young star models predict that higher mass stars will spin-down to their main sequence rotation track faster than low mass stars. This model could possibly explain our green outliers, where the A-component is rotating slower than the B, but not the blue outliers.
Furthermore, the latest type stars in our sample ($\sim0.4 M_{\odot}$) should converge after $\sim$1 Gyr. As field stars, the stars in our sample likely have ages older than 1 Gyr, so the anomalously fast rotators are unlikely to be young, low-mass stars that have yet to converge to a single evolutionary track.

A second alternative to the tidal spin-up hypothesis is the possibility that the affected systems are {\it too old} for gyrochronology models to accurately predict their ages. A recent study by \citet{vansaders2016} found unexpectedly rapid rotation in older main sequence stars, particularly those with a Rossby number above $Ro_{crit} = 2.16$. While all GKM stars reach $Ro_{crit} = 2.16$ on Gyr timescales, higher mass stars will hit this critical rotation limit earlier than lower mass stars, resulting in a possible rotation disparity where the A-component is rotating faster than expected compared to the B-component. However, as shown in Figure \ref{fig:rotation}, the outlier stars in our sample have Rossby numbers below $\sim$2. Further, this model cannot explain the green population, with anomalously fast rotating B-components.

\begin{figure*}[ht]
\centering
\includegraphics[width=0.45\textwidth]{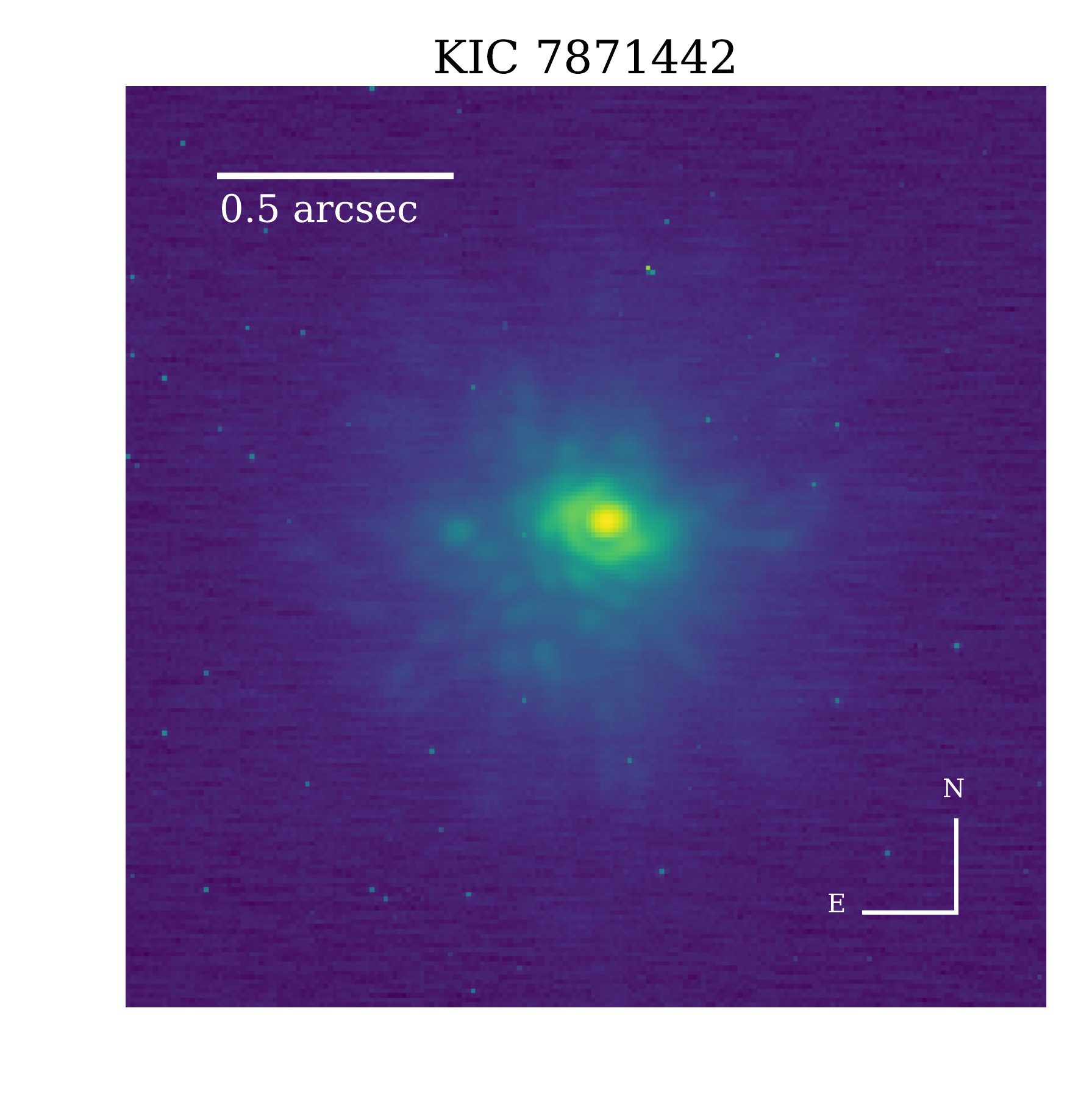}
\includegraphics[width=0.45\textwidth]{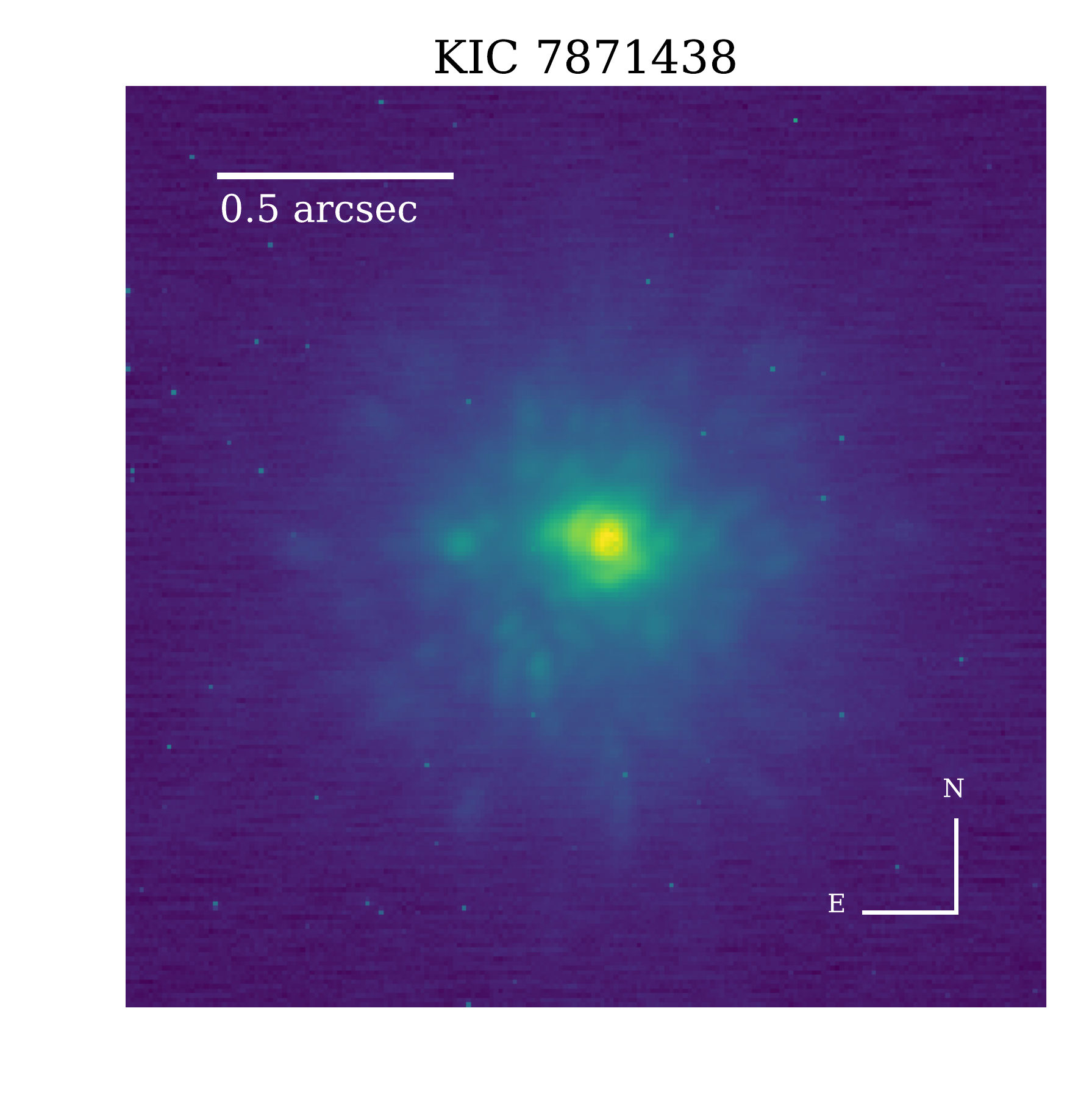}\\
\includegraphics[width=0.45\textwidth]{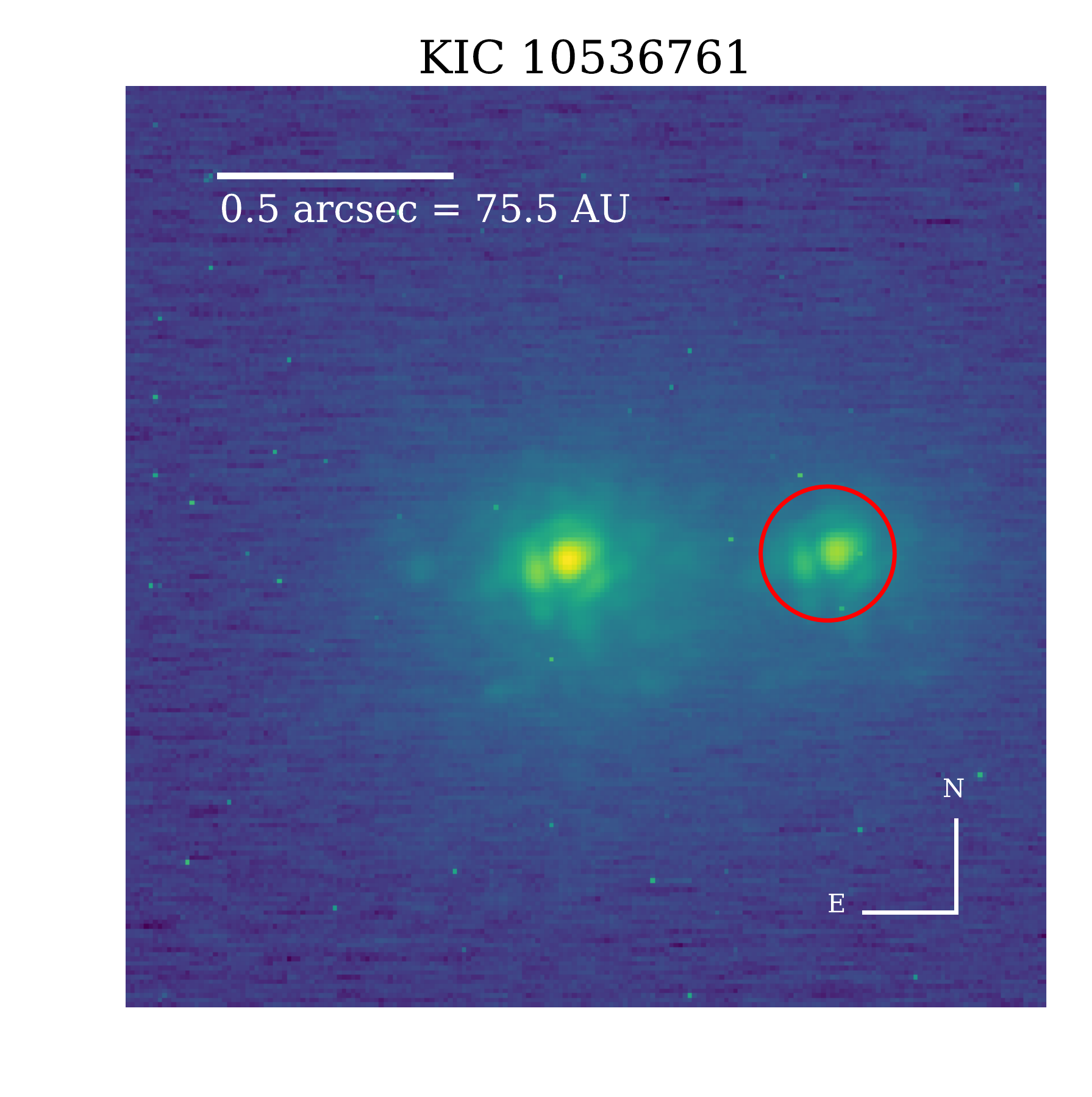}
\includegraphics[width=0.45\textwidth]{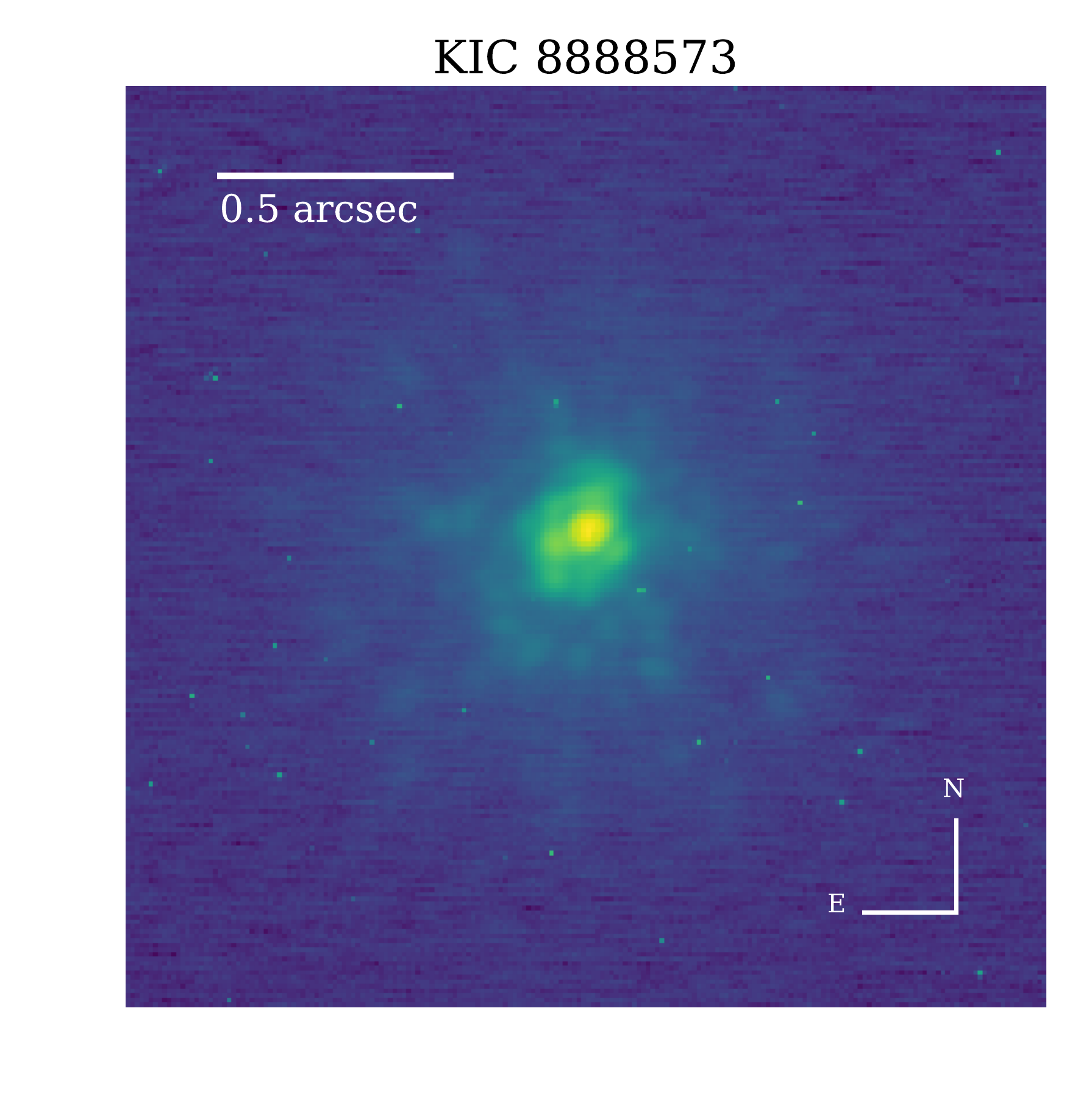}
\caption{Adaptive Optics images of both the A- and B-components for the two wide binary systems shown in Figure \ref{fig:lc}. A companion to KIC 10536761 is clearly detected (red circle) with a separation of 0.58 arcsec (87 AU), which may explain the multiple period signals apparent in the star's light curve. The small, nearby bright sources in the KIC 7871442 and KIC 7871438 images are speckles, not companion stars. \added{The speckles are non-astrophysical, as the separation varies with wavelength.}
}
\label{fig:ao}
\end{figure*}

As a proof-of-concept for our tidal spin-up hypothesis, we searched for previously unresolved tertiary companions using Adaptive Optics (AO) imaging. The observations of these stars is described in \S\ref{sec:AO}, and the resulting images are shown in Figure \ref{fig:ao}. Small, nearby bright points in the images are speckles from the AO imaging, which we confirmed as having an apparent separation that increased as a function of wavelength. We targeted the four objects whose light curves were shown in Figure \ref{fig:lc}, which were selected as having significant flare activity asymmetry between A- and B-components, as well as mass ratios near $q=1$, making them strong candidates for possessing a tertiary star. 

KIC 10536761, an anomalously fast rotating A-component, was found to have a low-contrast ($\sim$0.8 mag fainter) companion at 0.58'' angular separation (87 AU physical separation), as shown in Figure \ref{fig:ao}. Upon closer analysis, this companion is also evident in the original \Kepler light curve from Figure \ref{fig:lc}, which shows possibly two periodic signals.
We generated a Lomb-Scargle periodogram of the KIC 10536761 long cadence data, which revealed two potential rotation periods: the 5.65 day period reported by \citet{janes2017}, and a $\sim$1.1 day period that can be seen as a smaller amplitude modulation of the light curve. However, we cannot distinguish which star each period belongs to from the \Kepler data alone, wherein the system is unresolved. 
Note also: the presence of this low-contrast third body means KIC 10536761 is {\it not} strictly speaking the higher-mass component of this wide binary. Updated mass estimates for this newly discovered triple star system are needed.

However, no triple star was discovered in the AO data for KIC 7871438, the anomalously fast rotating B-component. This could be due to either the tidal spin-up occurring from a companion too close for our AO data to resolve, or from a companion that was previously ejected from the system. Alternatively, this may indicate that tidal spin-up is not the sole cause of these outlier systems. These systems are therefore good candidates to search for close companions. The occurrence rate of these outliers, particularly the ratio of blue to green systems, may put an independent constraint on tight, hierarchical triple star formation and evolution.

Beyond the presence of a third body, two other factors are important for the tidal spin-up of stars into the discrepant rotation periods we observe today. The first is the tightness of the interaction. The strength of the tidal force decreases as $F_{T} \propto 1/r^3$, meaning an interaction with a third companion would have to occur at very close orbital separations in order to exert a sufficient tidal spin-up on the affected star. \citet{lurie2017} recently found in a sample of unresolved \Kepler eclipsing binaries that tidal forces affected stellar rotation for systems with orbital periods shorter than 30 days, and typically led to orbit-spin synchronization for orbital periods shorter than 10 days. The faster rotators in the outlier pairs from our sample predominantly display $P_{rot} < 10$ days. However, at 87 AU physical separation, the inner binary of KIC 10536761 has an orbital period P $>>$ 30 days, which casts doubt on the potential for a tidal spin-up in this system in its current configuration. The inner binary may have been tighter earlier in the history of the system.

The second factor to consider is {\it when} the spin-up interaction occurs. The conclusion that tidal spin-up may be the mechanism responsible for pairs in the green and blue populations with mass ratios near $q=1$ relies on the assumption that both stars follow rotation evolution tracks such as those described in \citet{barnes&kim2010}. 
If tidal spin-up is the mechanism through which the observed asymmetry in $L_{fl}/L_{kp}$ is introduced, then the spin-up must have occurred after the stars converge to a single evolutionary track in period--mass space. Otherwise, the rotation periods of the spun-up star and its unperturbed counterpart would converge back to the same $P_{rot}$. As field stars, the stars in our sample are older than the convergence age of the latest stars in our sample, so the anomalously fast rotators are not simply young, low-mass stars that have yet to converge to a single evolutionary track. Furthermore, the critical Rossby number described in \citet{vansaders2016} may provide an upper limit on when the spin-up occurs. Cooler, less massive stars reach $Ro_{crit}$ at later ages, with the coolest stars ($T_{eff}$ = 5250 K) reaching $Ro_{crit} = 2.16$ at $\sim$ 6.5 Gyr, providing a constraint on when the spin-up can produce the observed effect on the star's age--rotation--activity profile.

\section{Summary}
\label{sec:summary}

By combining a list of known wide binaries in the \Kepler field to estimates of their relative flare luminosities, we have performed a unique comparison of flare activity in resolved, equal-mass binaries. From this sample, we made the following observations:

\begin{enumerate} 
\item{Flare activity comparison of 58 wide binaries found 48 systems consistent with current age--rotation and rotation--activity expectations.}
\item{Four systems defied the age--rotation--activity expectation for coeval, equal mass binaries. Two displayed anomalously elevated activity in the secondary component, while the other two displayed anomalously elevated flare activity in the primary component.}
\item{In the rotation--activity space, the outlier systems universally displayed faster rotation in the star with the higher flare rate, consistent with the \citet{davenport2016} flare catalog as well as with other tracers of magnetic activity.}
\item{Adaptive Optics follow-up observations revealed a previously unresolved tertiary in an equal-mass outlier system, reinforcing the possibility that the anomalous rotation--activity behavior could be due to dynamical effects such as tidal spin-up.}
\end{enumerate} 

While discerning the mechanism behind the unexpected rotation-activity of the outlier systems is beyond the scope of this paper, the fact that the majority of this sample follows the expected age--activity-rotation relation is a significant validation of flares as tracers of stellar magnetic activity. Larger samples of wide binaries with more robust filtering for contaminants may serve to more strongly validate this initial finding.

\acknowledgements

J.R.A.D. is supported by an NSF Astronomy and Astrophysics Postdoctoral Fellowship under award AST-1501418. C.B. acknowledges support from the Alfred P. Sloan Foundation.

This paper includes data collected by the Kepler mission. Funding for the Kepler mission is provided by the NASA Science Mission directorate. 

Some/all of the data presented in this paper were obtained from the Mikulski Archive for Space Telescopes (MAST). STScI is operated by the Association of Universities for Research in Astronomy, Inc., under NASA contract NAS5-26555. Support for MAST for non-HST data is provided by the NASA Office of Space Science via grant NNX09AF08G and by other grants and contracts. 

Some of the data presented herein were obtained at the W. M. Keck Observatory, which is operated as a scientific partnership among the California Institute of Technology, the University of California and the National Aeronautics and Space Administration. The Observatory was made possible by the generous financial support of the W. M. Keck Foundation. 

The authors wish to recognize and acknowledge the very significant cultural role and reverence that the summit of Maunakea has always had within the indigenous Hawaiian community.  We are most fortunate to have the opportunity to conduct observations from this mountain.

\facility{Kepler, Keck:II (NIRC2)}

\startlongtable
\begin{deluxetable}{cccccc}
\tablecaption{Properties of the final 116 star sample. Rotation periods are from \citet{janes2017} and the masses are from \citet{davenport2016}. Note that some stars do not have a measured \textit{g-i} in the KIC.\label{table}}
\tablehead{\colhead{KIC} & \colhead{Binary \#} & \colhead{g-i} & \colhead{$P_{rot}$ [days]} & \colhead{$L_{fl}/L_{kp}$} & \colhead{Mass $[M_{\odot}]$}}
\startdata
7090654 & 1 & 0.43 & 1.98 & 3.4e-05 & 1.163 \\
7090649 & 1 & 0.78 & 8.27 & 1.15e-06 & 0.923 \\
7659402 & 2 & 2.12 & 19.66 & 3.04e-06 & 0.607 \\
7659417 & 2 & 2.27 & 16.31 & 1.38e-05 & 0.571 \\
7582687 & 3 & 0.72 & 14.07 & 3.25e-07 & 0.949 \\
7582691 & 3 & 1.89 & 24.54 & 1.16e-07 & 0.655 \\
8006740 & 4 & \ldots & 22.84 & 1.63e-07 & 0.703 \\
8143903 & 4 & 2.40 & 41.97 & 8.1e-07 & 0.537 \\
5936797 & 5 & 0.67 & 25.72 & 5.5e-06 & 0.980 \\
5936811 & 5 & 1.09 & 34.12 & 1.19e-05 & 0.808 \\
7093953 & 6 & 0.58 & 24.78 & 7.22e-07 & 1.036 \\
7093968 & 6 & 1.16 & 36.72 & 7.96e-07 & 0.788 \\
10255692 & 7 & 2.18 & 21.58 & 4.75e-07 & 0.588 \\
10255689 & 7 & 2.55 & 31.82 & 1.12e-06 & 0.483 \\
7871442 & 8 & 2.04 & 17.45 & 1.06e-07 & 0.625 \\
7871438 & 8 & 2.08 & 2.85 & 0.000286 & 0.594 \\
11069662 & 9 & 0.57 & 30.96 & 4.75e-07 & 1.055 \\
11069655 & 9 & 0.59 & 3.51 & 3.63e-07 & 1.023 \\
10388283 & 10 & 2.50 & 39.01 & 1.46e-05 & 0.494 \\
10388259 & 10 & 2.61 & 48.28 & 6.58e-06 & 0.465 \\
10518551 & 11 & 0.45 & 20.84 & 3.19e-08 & 1.122 \\
10518563 & 11 & 1.19 & 28.69 & 1.29e-06 & 0.782 \\
9139163 & 12 & 0.34 & 0.61 & 3.29e-07 & 1.263 \\
9139151 & 12 & 0.45 & 12.22 & 1.58e-07 & 1.155 \\
4995565 & 13 & 0.72 & 15.44 & 1.29e-08 & 0.941 \\
4995581 & 13 & 2.45 & 41.45 & 2.7e-06 & 0.516 \\
4043389 & 14 & 2.39 & 38.84 & 9.64e-07 & 0.544 \\
4142913 & 14 & \ldots & 37.84 & 3.31e-07 & 0.361 \\
6678383 & 15 & 0.51 & 11.42 & 7.3e-07 & 1.080 \\
6678367 & 15 & 0.86 & 17.41 & 1.09e-06 & 0.884 \\
9579208 & 16 & 0.38 & 5.5 & 7.04e-08 & 1.213 \\
9579191 & 16 & 0.65 & 14.66 & 2.71e-07 & 0.991 \\
7432575 & 17 & 1.69 & 15.62 & 1.6e-06 & 0.692 \\
7432573 & 17 & 4.06 & 19.66 & 1.87e-06 & 0.601 \\
9762519 & 18 & 0.72 & 7.73 & 0.000103 & 0.942 \\
9762514 & 18 & 1.17 & 16.08 & 6.03e-06 & 0.779 \\
7596937 & 19 & 0.62 & 10.38 & 3.97e-07 & 1.016 \\
7596922 & 19 & 1.11 & 19.31 & 1.3e-06 & 0.799 \\
12456757 & 20 & 0.61 & 27.0 & 2.51e-07 & 1.036 \\
10529126 & 20 & 0.73 & 10.29 & 2.87e-06 & 0.945 \\
7676737 & 21 & 1.74 & 0.72 & 3.98e-07 & 0.673 \\
11709022 & 21 & \ldots & 4.52 & 0.000511 & 0.665 \\
12507868 & 22 & 0.49 & 6.77 & 5.37e-08 & 1.083 \\
7676799 & 22 & 2.65 & 43.32 & 4.04e-06 & 0.433 \\
7885518 & 23 & 0.62 & 24.82 & 6.01e-07 & 0.990 \\
12507882 & 23 & 1.96 & 19.25 & 6.48e-07 & 0.645 \\
11861593 & 24 & 1.26 & 53.99 & 8.27e-07 & 0.764 \\
11241109 & 24 & 2.51 & 39.13 & 1.81e-06 & 0.523 \\
2442687 & 25 & \ldots & 27.08 & 1.67e-07 & 1.073 \\
7750144 & 25 & 0.77 & 3.63 & 3.17e-06 & 0.940 \\
7118431 & 26 & 1.19 & 39.07 & 1.14e-07 & 0.781 \\
3955963 & 26 & 2.01 & 35.21 & 9.09e-07 & 0.639 \\
7118479 & 27 & 0.36 & 42.87 & 9e-07 & 1.022 \\
8098178 & 27 & 2.24 & 18.89 & 0.000148 & 0.570 \\
2992956 & 28 & 0.68 & 0.99 & 9.53e-08 & 0.999 \\
8098181 & 28 & 1.18 & 0.98 & 0.0012 & 0.760 \\
2992960 & 29 & 0.74 & 22.14 & 3.85e-10 & 0.962 \\
7364380 & 29 & 1.06 & 24.51 & 1.48e-07 & 0.818 \\
10275409 & 30 & 0.60 & 48.65 & 2.11e-07 & 1.019 \\
7364389 & 30 & 2.40 & 45.32 & 1.94e-06 & 0.547 \\
10536753 & 31 & \ldots & 7.2 & 0.000156 & 1.178 \\
10275420 & 31 & 0.76 & 9.89 & 1.57e-06 & 0.932 \\
10536761 & 32 & 2.41 & 0.79 & 0.000129 & 0.524 \\
8888573 & 32 & 2.69 & 11.22 & 1.41e-07 & 0.427 \\
4931390 & 33 & 0.30 & 31.37 & 3.98e-07 & 1.253 \\
4931385 & 33 & 1.96 & 16.33 & 5.83e-07 & 0.646 \\
8565874 & 34 & 0.57 & 12.27 & 2.93e-05 & 1.042 \\
8565877 & 34 & 0.62 & 7.62 & 3.08e-07 & 0.987 \\
9897318 & 35 & 0.96 & 0.0 & 1.3e-06 & 0.859 \\
9897328 & 35 & 1.03 & 11.32 & 1.97e-06 & 0.830 \\
12214504 & 36 & 0.70 & 12.77 & 1.74e-06 & 0.969 \\
12214492 & 36 & 0.94 & 3.86 & 3.77e-06 & 0.856 \\
12068975 & 37 & 0.42 & 16.28 & 6.33e-07 & 1.147 \\
12068971 & 37 & 1.27 & 10.38 & 9.06e-07 & 0.754 \\
11515925 & 38 & 1.12 & 17.97 & 3.65e-07 & 0.702 \\
11515931 & 38 & 2.09 & 16.43 & 1.3e-06 & 0.614 \\
5202445 & 39 & 1.40 & 20.53 & 3.51e-06 & 0.703 \\
5202421 & 39 & 1.90 & 17.25 & 2.06e-06 & 0.659 \\
10612448 & 40 & 2.07 & 14.19 & 6.18e-07 & 0.624 \\
10612424 & 40 & 2.33 & 21.08 & 5.1e-07 & 0.567 \\
4484238 & 41 & \ldots & 20.38 & 3.9e-07 & 1.061 \\
4386086 & 41 & 0.69 & 36.54 & 5.97e-07 & 1.000 \\
12317678 & 42 & 0.29 & 9.62 & 1.54e-07 & 1.267 \\
12218888 & 42 & 0.48 & 24.77 & 2.24e-09 & 1.121 \\
8248671 & 43 & 0.52 & 6.54 & 3.78e-07 & 1.101 \\
8248626 & 43 & 0.83 & 5.45 & 3.78e-07 & 0.897 \\
12024098 & 44 & \ldots & 12.75 & 1.48e-09 & 1.445 \\
12024088 & 44 & 1.46 & 13.3 & 1.6e-06 & 0.706 \\
11724888 & 45 & 0.83 & 32.55 & 4.22e-06 & 0.931 \\
11724885 & 45 & 0.85 & 31.12 & 8.68e-07 & 0.920 \\
6225718 & 46 & \ldots & 16.34 & 1.24e-07 & 1.096 \\
6225816 & 46 & \ldots & 30.85 & 0.000438 & 0.722 \\
11876220 & 47 & 2.36 & 7.32 & 4.54e-05 & 0.557 \\
11876227 & 47 & 2.76 & 2.17 & 0.000657 & 0.420 \\
10616124 & 48 & -0.06 & 1.47 & 1.25e-05 & 1.724 \\
10616138 & 48 & 0.01 & 1.46 & 8.19e-06 & 1.591 \\
8184081 & 49 & 0.54 & 2.82 & 1.69e-06 & 1.065 \\
8184075 & 49 & 0.91 & 0.74 & 7.75e-07 & 0.866 \\
10355856 & 50 & 0.31 & 13.18 & 3.05e-07 & 1.255 \\
10355809 & 50 & 0.70 & 13.51 & 0.000272 & 0.703 \\
5211089 & 51 & 1.69 & 4.49 & 2.41e-06 & 0.699 \\
5211083 & 51 & 1.89 & 1.52 & 1.08e-06 & 0.658 \\
11098013 & 52 & 0.38 & 41.64 & 1.22e-07 & 1.199 \\
11098004 & 52 & 0.47 & 39.5 & 8.54e-07 & 1.122 \\
6545403 & 53 & \ldots & 5.38 & 5.61e-07 & 0.833 \\
6545415 & 53 & 2.12 & 5.27 & 5.86e-07 & 0.605 \\
4864392 & 54 & 0.94 & 22.16 & 3.54e-07 & 0.853 \\
4864391 & 54 & 2.02 & 18.2 & 8.01e-06 & 0.636 \\
10230145 & 55 & 0.97 & 19.82 & 1.73e-08 & 0.851 \\
10296031 & 55 & 2.37 & 26.84 & 3.45e-06 & 0.516 \\
10622511 & 56 & 2.27 & 44.01 & 2.43e-06 & 0.580 \\
10557342 & 56 & 2.64 & 16.09 & 4.53e-07 & 0.463 \\
8909853 & 57 & 2.30 & 37.92 & 8.34e-06 & 0.562 \\
8909876 & 57 & 2.39 & 17.23 & 3.11e-06 & 0.539 \\
10164867 & 58 & 0.32 & 38.86 & 1.35e-07 & 1.221 \\
10164839 & 58 & 0.33 & 53.37 & 1.9e-07 & 1.211 \\
\enddata
\end{deluxetable}
 \(\)
\end{document}